\documentclass[twocolumn,showpacs,preprintnumbers,amsmath,amssymb,superscriptaddress]{revtex4}
\usepackage[toc,page]{appendix}
\usepackage{graphicx}
\usepackage{dcolumn}
\usepackage{bm}
\usepackage{color}
\usepackage{amsbsy}

\begin{document}

\title{Photon statistics of the light transmitted and reflected by a two-dimensional atomic array}

\author{Daniel Cano}
\affiliation{Institute of Physical and Information Technologies, Spanish National Research Council (CSIC), 28006 Madrid, Spain}

\begin{abstract}

This work proposes to investigate the photon statistics of the light transmitted and reflected by a two-dimensional array of interacting atoms. The reflected beam is characterized by photon antibunching. On the other hand, in the transmitted beam the indistinguishability between the driving laser photons and the photons re-emitted by the atoms results in photon bunching. The overlap between the driving and scattered fields is enhanced by the cooperative optical response of the atomic array. In the examples used in this paper, up to 25\% of the transmitted photons are grouped in pairs. The simulations are carried out using the stochastic method of quantum trajectories.

\end{abstract}

\maketitle

\section{Introduction} \label{sec:introduction}

Optical metasurfaces based on sub-wavelength arrays of light scatterers offer the possibility of modifying the properties of light over a scale much smaller than the free-space wavelength \cite{Yu:14,Chen:16}. While most applications so far operate in the classical light regime, there is growing interest in creating optical metasurfaces for quantum technologies \cite{Stav:18,Wang:18,Lyons:19}. Recently, optical metasurfaces were realized using two-dimensional lattices of periodically spaced atoms \cite{Rui:20}, with great potential for quantum information processing \cite{Bekenstein:20}. In these systems, the photon-mediated dipole-dipole interactions cause the atoms to behave as a collective rather than independently. This cooperative optical response leads to a number of intriguing properties. One of the most remarkable effects is that, for certain lattice spacings, a two-dimensional atomic array can act as a mirror reflecting most of the electromagnetic energy of a Gaussian beam \cite{Bettles:15,Bettles:16,Shahmoon:17,Rui:20,Yoo:20}. In addition, this kind of atomic monolayer exhibits a variety of interesting non-linear quantum optical effects, such as optical phase transitions \cite{Parmee:20}, bistable optical transmission \cite{Parmee:21} and light-induced spin-spin correlations \cite{Bettles:20}.

The use of optical metasurfaces in quantum technologies still requires a better understanding of their cooperative optical behavior \cite{Brandes:05,Guerin:17}. Previous works on cooperative optical phenomena have focused mainly on spontaneous emission effects such as subradiance, superradiance, directional scattering and subradiant excitations \cite{Svidzinsky:10,Feng:13,Feng:14,Facchinetti:16,Arruda:20,Zhang:20,Alaee:20,Petrosyan:21,Gulfam:18,Masson:20,Bhatti:15,Liberal:19,Williamson:20,Holzinger:21}. These works assume that the photons emitted by the atoms can be distinguished from the photons of the driving laser field. However, this assumption is not always valid for metasurfaces and two-dimensional atomic arrays (see Fig. \ref{fig:Figure1}). Indeed, for certain lattice spacings, the field scattered by a two-dimensional atomic array has a spatial mode profile very similar to that of a focused Gaussian beam \cite{Bettles:16}. This implies that a large fraction of the photons collectively emitted by the atoms is indistinguishable from the photons of the driving laser field. This can, in principle, give rise to quantum interference and entanglement, with important implications for photon statistics.

The indistinguishability between the driving laser photons and the photons reemitted into the laser beam mode has an important effect on photon statistics because it is closely related to photon absorption and stimulated emission \cite{Einstein:16}. Indeed, absorption and stimulated emission require an overlap between the spatial mode profiles of the driving laser field and the dipole field of the atom. Interferences between both fields change over time from destructive to constructive during a Rabi cycle. When the interference is destructive (constructive), there is absorption (stimulated emission) \cite{Cray:82}. An interesting experiment would be to place a photodetector in the transmitted beam and to measure the photon statistics associated with absorption and stimulated emission. Unfortunately, the realization of this experiment with a single atom involves many technical complications that have not yet been solved, as it requires focusing the laser beam tightly to the same size as the absorption cross section of the atom \cite{Tey:08,Tey:09}. This paper will show that two-dimensional atomic arrays can facilitate a similar experiment with less demanding focusing conditions than with a single atom. This is possible thanks to the cooperative optical response of the atomic array, which leads to an increased optical cross section.

This paper simulates the photon statistics of the light transmitted and reflected by a two-dimensional array of interacting atoms in which the spatial mode profiles of the driving and scattered fields overlap (see Fig. \ref{fig:Figure1}). For this purpose, this paper reformulates the master equation of collective spontaneous emission \cite{Carmichael:00,Clemens:03,Clemens:04} in order to consider the indistinguishability between the driving laser photons and the scattered photons in the transmitted beam. A quantum description of the driving laser is incorporated into the interaction Hamiltonian using the theoretical framework developed in previous works \cite{Carmichael:93,Gardiner:94,Nha:05,Noh:08,Zhang:18}. Simulations are carried out using a quantum trajectory algorithm \cite{Dalibard:92,Molmer:96,Carmichael:00,Lambropoulos:06}. The interference between the driving and scattered fields results in the generation of photon bunching in the transmitted beam. Interestingly, the photon statistics in the atomic array is very similar to that caused by the processes of absorption and stimulated emission in a single atom. The recent developments of two-dimensional atomic arrays with an enhanced optical cross section \cite{Rui:20} represent a great opportunity to observe these effects with less demanding focusing requirements than in a single-atom system.

\section{Physical description of the system} \label{sec:Description}

\begin{figure}
\centerline{\scalebox{0.43}{\includegraphics{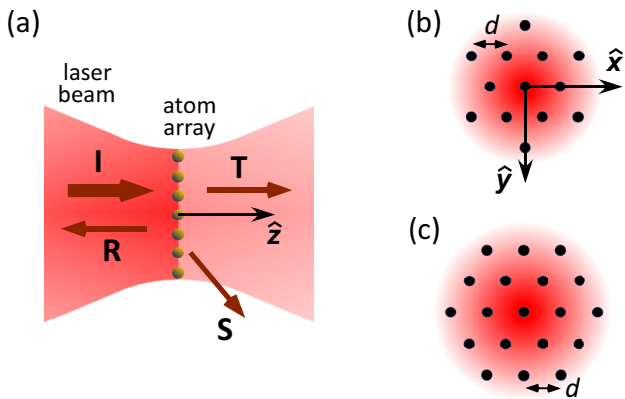}}}
\caption{Sketch of the system. (a) A two-dimensional atom array in the $xy$ plane is illuminated by a resonant laser beam propagating in the $z$ direction. The incoming beam is focused on the plane of the array. The incoming energy (I) splits into a reflected beam (R) and a transmitted beam (T). A few photons are scattered (S) in all directions. (b) Hexagonal geometry with $N$=13. The atomic separation is $d$. (c) Hexagonal geometry with $N$=19.} \label{fig:Figure1}
\end{figure}

The system consists of a two-dimensional array of identical two-level atoms located at positions $\bm r_n = (x_n,y_n)$, where $n=1,\dots,N$ (see Fig. \ref{fig:Figure1}). The atoms are illuminated by a resonant laser beam propagating in the $z$ direction. The beam is focused on the plane of the atoms and has a Gaussian intensity profile. The lattice is chosen to be hexagonal with sub-wavelength spacing $d$ in order to maximize the overlap between the driving and scattered fields [see Figs. \ref{fig:Figure1}(b) and \ref{fig:Figure1}(c)]. A homogeneous magnetic field is applied to keep the atomic dipole moments aligned with the laser polarization.

The beam field is in a monochromatic coherent state $|\alpha\rangle$, whose average number of photons per unit time is $|\alpha|^2$. The complex number $\alpha$ is given by $\alpha=\mathcal{E}/\mathcal{E}_1$, where $\mathcal{E}$ is the classical electric field and $\mathcal{E}_1 = \sqrt{(4 \pi \hbar)/(\epsilon_0 \lambda A_{\text{beam}})}$ is the electric field of one photon at the center of the focus. Here, $A_{\text{beam}}=\pi w_0^2 /2$ is the effective cross section of the beam, where $w_0$ is the beam waist.

The Rabi frequency at the center of the focus is given by $\Omega = -\mathcal{E} p_{eg} / \hbar$, where $p_{eg} = \sqrt{(3 \epsilon_0 \hbar \lambda^3 \Gamma)/(8 \pi^2)}$ is the dipole moment of the atomic transition. Here, $\lambda$ is the photon wavelength and $\Gamma$ is the single-atom decay rate. The number $\alpha$ can be calculated directly from $\Omega$ using $\alpha = -(\hbar \Omega)/(p_{eg}\mathcal{E}_1)$.

\section{Optimal overlap between the driving and scattered fields} \label{sec:Optimal parameters}

Simulations are carried out with a lattice geometry that optimizes the overlap between the spatial mode profiles of the driving and scattered fields. A hexagonal lattice with subwavelength spacing is a good choice for this purpose because it scatters light with a spatial mode profile very similar to a Gaussian beam focused on the plane of the atoms \cite{Bettles:16}. In such a hexagonal array the off-axis scattering is suppressed. In addition, the laser beam has to be focused so that its size is similar to or smaller than the optical cross section of the atomic array. The optimal values of $w_0$ and $d$ are found using the methods of Refs. \cite{Bettles:15,Bettles:16} (see Fig. \ref{fig:Figure1}). This geometry is optimized only for a resonant laser.

One of the properties of optimal overlap between the driving and scattered fields is that, in the low-laser-intensity limit, the atomic array behaves like a mirror because both fields interfere destructively with each other. In this case, the reflected light maintains the Poissonian character of the incoming laser beam. The low-laser-intensity limit was studied in detail in Refs. \cite{Bettles:15,Bettles:16,Shahmoon:17,Rui:20,Yoo:20}. The opposite happens in the high laser intensity limit ($\Omega \gg \Gamma$). In this case, the atomic transition saturates, allowing most of the light to be transmitted. Then, the transmitted beam retains the Poissonian character of the incoming beam.

The most interesting situation occurs for intermediate intensities, in which a significant fraction, although not the majority, of the incoming photons is transmitted. This is the situation in which the atomic array modifies the Poissonian statistics of the laser field to the greatest extent, and this is the case studied in this paper.

\section{Theoretical model}

\subsection{Quantum trajectory algorithm} \label{sec:Algorithm}

The stochastic method of quantum trajectories offers a natural way to study photon statistics in few-atom systems \cite{Dalibard:92,Molmer:96,Carmichael:00,Lambropoulos:06}. The numerical algorithm used in this work assumes that photons are detected by means of a set of imaginary photodetectors covering the whole surface of a sphere located at far-field distances. The quantum trajectory algorithm simulates the time evolution of the atomic states as well as the photon-counting records in the photodetectors. The numerical method consists of calculating a series of unnormalized stochastic wave functions $|\psi_j (t)\rangle$ $\left( j=1,\dots,J \right)$, each of which represents the quantum trajectory of a thought experiment \cite{Dalibard:92,Molmer:96,Carmichael:00,Lambropoulos:06}. Every wavefunction $|\psi_j (t)\rangle$ evolves accordingly to the Sch\"{o}dinger equation
\begin{equation}
\frac{d}{dt} |\psi_j (t)\rangle = -\frac{i}{\hbar}{\cal H} |\psi_j (t) \rangle, \label{Eq:Schroedinger}
\end{equation}
using the non-Hermitian Hamiltonian
\begin{equation}
{\cal H}={\cal H}_{AF} + {\cal H}_{AA} - i \frac{\hbar}{2} \sum_{q=1}^{Q} \sum_{s} P^\dag_{qs} P_{qs}, \label{Eq:H_total}
\end{equation}
where ${\cal H}_{AF}$ is the atom-field interaction Hamiltonian, ${\cal H}_{AA}$ is the atom-atom interaction Hamiltonian, and $P_{qs}$ is the jump operator associated with the detection of a photon on detector $q$ with polarization $s$. The mathematical formulas of these operators are described in Sec. \ref{sec:Formulation}.

Initially, at time $t = 0$, $|\psi_j (t)\rangle$ is assumed to be normalized. Since ${\cal H}$ is non-Hermitian, the norm of $|\psi_j (t)\rangle$ decreases with time, reflecting the fact that the system is not isolated. A photon is detected when the squared norm of the wavefunction has decreased to $|\langle \psi_j | \psi_j \rangle |^2 = r_1$,  where $r_1$ is a random number between 0 and 1. The detector on which the photon is detected is determined from the Monte Carlo probabilities $\langle \psi_j(t) | P_{qs}^\dag P_{qs} |\psi_j (t) \rangle$. At the instant of time immediately after photon detection, the wave function $|\psi_j (t)\rangle$ is projected using the corresponding jump operator,
\begin{equation}
|\psi_j (t) \rangle   \longrightarrow  \frac{P_{qs} |\psi_j (t) \rangle}{ \sqrt{\langle \psi_j(t) | P_{qs}^\dag P_{qs} |\psi_j (t) \rangle }}.  \label{Eq:projection}
\end{equation}
This projection produces a sudden change in the quantum state of the system, the so-called quantum jump. Then, the computer generates another random number, $r_2$. The second photon is detected when the norm of the wavefunction has decreased to $|\langle \psi_j | \psi_j \rangle |^2 = r_2$. The process continues until the desired number of photon counting records has been obtained. For each trajectory $j$, we have the series of times at which the photons are detected $\{t_{j1},t_{j2},t_{j3},\dots \}$ and the corresponding series of detectors $\{q_{j1},q_{j2},q_{j3},\dots \}$.

\subsection{Mathematical formulas of the quantum operators} \label{sec:Formulation}

The simulations take into account the two following facts: (i) The decay rates and the transition energies are modified by the dipole-dipole interactions between atoms. (ii) The electromagnetic far field is the sum of the driving laser field and the quantum field emitted by the atoms. The simulations use the master equation of collective spontaneous emission \cite{Carmichael:00,Clemens:03,Clemens:04} with a quantum description of the laser field incorporated into it. For this purpose, the laser field is described using the theoretical framework of Refs. \cite{Carmichael:93,Gardiner:94,Nha:05,Noh:08,Zhang:18}.

\subsubsection{Atom-field interaction Hamiltonian} \label{sec:H_AF}

The atom-field interaction Hamiltonian in the rotating-wave approximation is
\begin{equation}
{\cal H}_{AF} = \frac{\hbar}{2} g \sum_{n=1}^{N} \left( f_n a \sigma_n^\dag + f_n^\ast a^\dagger \sigma_n \right),
\label{Eq:H_AF}
\end{equation}
where $g$ is the atom-field coupling constant at the center of the beam focus, $f_n = \exp{\left[ -(x_n^2+y_n^2)/w_0^2 \right]}$ accounts for the spatial variations of the field, $a$ is the annihilation operator of the field mode, and $\sigma_n \equiv |g_n\rangle \langle e_n|$ is the operator that lowers the state of atom $n$. The coupling constant is
\begin{equation}
g=-\frac{\mathcal{E}_1 p_{eg}}{2 \hbar}. \label{Eq:g}
\end{equation}

It is very practical to express $g$ as a function of $\Gamma$. From Sec. \ref{sec:Description} and Eq. \ref{Eq:g}, we find
\begin{equation}
g= \sqrt{\vartheta \Gamma}, \label{Eq:g_bis}
\end{equation}
where the parameter $\vartheta=A_{\text{atom}}/(4 A_{\text{beam}})$ represents the overlap between the laser field mode and the dipole-field mode of an atom located at the focus center.

The Hamiltonian ${\cal H}_{AF}$ in Eq. \ref{Eq:H_AF} has the same form as an atom-cavity interaction Hamiltonian \cite{Lambropoulos:06}. This result may be surprising since photon reemission into the propagating laser-field mode is not a reversible process. The irreversibility of this process is actually taken into account by adding the product of jump operators $P^\dag_{qs} P_{qs}$ in the total Hamiltonian ${\cal H}$ in Eq. \ref{Eq:Schroedinger}. The product $P^\dag_{qs} P_{qs}$ contains additional terms $a \sigma_n^\dag$ that account for photon scattering. The demonstration that ${\cal H}_{AF}$ actually represents the atom-field interaction Hamiltonian can be found in Appendixes \ref{sec:Master} and \ref{sec:Comparacion}.

\subsubsection{Atom-atom interaction Hamiltonian} \label{sec:H_AA}

Atoms interact with each other via photon exchange. The atom-atom interaction Hamiltonian is
\begin{equation}
{\cal H}_{AA} = \hbar \sum_{n \neq m=1}^{N} \Delta_{nm} \sigma_n^\dag \sigma_m,
\label{Eq:H_AA}
\end{equation}
where the energy shift $\Delta_{nm}$ can be derived from the dipole-dipole potential \cite{Carmichael:00},
\begin{eqnarray}
\Delta_{nm}  & = & \frac{3 \Gamma}{4} \left[ - \left( 1-|\bm{\hat{u}} \cdot \bm{\hat{r}_{nm}}|^2\right) \frac{\cos \xi_{nm}}{\xi_{nm}} \right. \nonumber \\
 & + & \left. \left( 1-3 |\bm{\hat{u}} \cdot \bm{\hat{r}_{nm}}|^2\right) \left(\frac{\sin \xi_{nm}}{\xi_{nm}^2}+\frac{\cos \xi_{nm}}{\xi_{nm}^3} \right) \right], \label{Eq:Delta1}
\end{eqnarray}
where $\bm{\hat{u}}$ is the polarization unit vector of the atomic dipole moment, $\bm{\hat{r}_{nm}}$ is the unit vector in the direction of $\bm{r_{nm}}=\bm{r_{n}}-\bm{r_{m}}$, and $\xi_{nm}\equiv 2 \pi r_{nm}/\lambda$.

\subsubsection{Jump projectors for $N=1$} \label{sec:Method_OneAtom}

The single-atom system can be simulated with only two jump operators: one for the direction of the laser beam (forward scattering) and one for all other directions (backward and sideways scattering). These operators are respectively \cite{Carmichael:93,Gardiner:94,Nha:05,Noh:08,Zhang:18}
\begin{equation}
P_f = - i a + \sqrt{\vartheta \Gamma} \sigma \hspace{2mm}, \hspace{5mm} P_b = \sqrt{(1-\vartheta) \Gamma } \sigma , \label{Eq:Jump_operators_N1}
\end{equation}
where $\vartheta \Gamma$ is the decay rate into the laser beam mode and $(1-\vartheta) \Gamma$ is the decay rate in all other directions. Notice that $P_f$ is the sum of two terms, which represent the detection of a laser photon and the detection of a scattered photon. To correctly consider the phases of the fields, the Gouy phase $-i$ must be written in front of $a$ \cite{Jackson:67}. The next section will show how to extend this formalism to many-atom arrays.

\subsubsection{Jump operators for $N>1$} \label{sec:Method_NAtoms}

\begin{figure}
\centerline{\scalebox{0.6}{\includegraphics{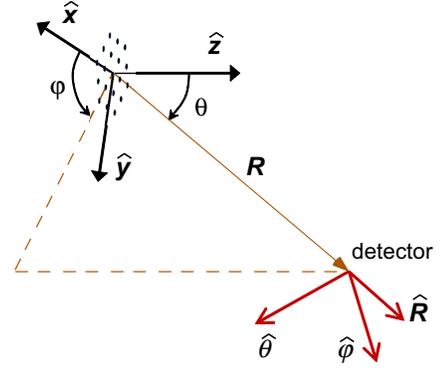}}}
\caption{The Cartesian vector basis $\{\bm{\hat{x}} , \bm{\hat{y}} , \bm{\hat{z}}\}$ is used in the plane of the atom array. The basis $B \equiv \{\bm{\hat{R}} , \bm{\hat{\theta}} , \bm{\hat{\phi}} \}$ of the orthogonal unit vectors in the directions of increasing spherical coordinates is used at the positions of the detectors. Here, $\theta$ is the polar angle with respect to the direction $z$, and $\phi$ is the azimuthal angle.} \label{fig:Figure_axes}
\end{figure}

The jump operators $P_{qs}$ are defined such that $\langle \psi_j(t) | P_{qs}^\dag P_{qs} |\psi_j (t) \rangle$ is the photon flux with polarization $s$, at time $t$, registered by photodetector $q$. We can then write \cite{Jackson:67}
\begin{equation}
P_{qs}=\sqrt{\frac{2c\epsilon_0}{\hbar \omega}} \bm E^{(+)}_T(R,\theta_q,\phi_q) \cdot \bm{\hat{s}}(\theta_q,\phi_q),
\label{Eq:jump_operators_1}
\end{equation}
where $\bm E^{(+)}_T(R,\theta_q,\phi_q)$ is the electric-field operator at photodetector $q$ and $\bm{\hat{s}}$ is the polarization unit vector in the direction of polarization $s$. The total number of photodetectors $Q$ must be high enough that the directional emission pattern is correctly simulated and the electromagnetic energy is conserved. In typical simulations, $Q \gtrsim 700$. The photodetectors cover the whole surface of a sphere of radius $R$ at far-field distances. The position of each photodetector $q$ is expressed in spherical coordinates $(R, \theta_q,\phi_q)$. The two polarization states of the far field, $\bm{\hat{s}}=\bm{\hat{\theta}},\bm{\hat{\phi}}$, are the orthogonal unit vectors in the directions of increasing spherical coordinates, $\theta$ and $\phi$, respectively (see Fig. \ref{fig:Figure_axes}).

The electric-field operator is $\bm E^{(+)}_T(R,\theta,\phi)=\bm E^{(+)}_D(R,\theta,\phi)+\bm E^{(+)}_L(R,\theta,\phi)$, where $\bm E^{(+)}_D(R,\theta,\phi)$ is the dipole field emitted by the atoms and $\bm E^{(+)}_L(R,\theta,\phi)$ is the laser field. The former is given by \cite{Jackson:67}
\begin{equation}
\bm E^{(+)}_D(R,\theta,\phi) = \frac{k^2 p_{eg}}{4\pi\epsilon_0} \bm{u}_D(\theta,\phi) \frac{e^{ikR}}{R}\tilde{\sigma}(\theta,\phi),
\label{Eq:E_D}
\end{equation}
where $k$ is the wave number,
\begin{equation}
\tilde{\sigma}(\theta,\phi) = \sum_{n=1}^{N} e^{-ik \bm{\hat{r}_n} \cdot  \bm{\hat{R}}} \sigma_n
\label{Eq:collective_sigma}
\end{equation}
is the collective atomic operator, $\bm{\hat{r}_n}$ is the unit vector in the direction of $\bm{r_n}$, $\bm{\hat{R}}=(\sin \theta \cos \phi, \cos \theta \cos \phi, -\sin \phi)$, and $\bm{u}_D(\theta,\phi) = T(\theta,\phi) \left[  ( \bm{\hat{R}} \times \bm{\hat{u}} ) \times \bm{\hat{R}} \right]$. Here, $T(\theta,\phi)$ is the change-of-basis matrix from $\{\bm{\hat{x}} , \bm{\hat{y}} , \bm{\hat{z}} \}$ to $\{\bm{\hat{R}} , \bm{\hat{\theta}} , \bm{\hat{\phi}} \}$, which is given by
\begin{equation*} \label{Eq:HS}
T(\theta,\phi) =
\begin{pmatrix}
\sin \theta \cos \phi & \sin \theta \sin \phi & \cos \theta \\
\cos \theta \cos \phi & \cos \theta \sin \phi & -\sin \theta \\
-\sin \phi & \cos \phi & 0
\end{pmatrix}.
\end{equation*}

The laser field operator is,
\begin{equation}
\bm E^{(+)}_L(R,\theta,\phi) = -i \mathcal{E}_1 \bm{F}(\theta,\phi) \frac{e^{ikR}}{R} a,
\label{Eq:E_L}
\end{equation}
with
\begin{equation}
\bm F(\theta,\phi) \simeq  \bm{\hat{u}}_L(\phi) \frac{z_R}{\cos \theta} \exp{\left[-\left( \frac{\pi w_0}{\lambda} \tan \theta \right)^2\right]},
\label{Eq:F}
\end{equation}
where $\bm{\hat{u}}_L(\phi) = T(\theta=0,\phi) \bm{\hat{u}}$, and $z_R=\pi w_0^2/\lambda$ is the Rayleigh length. Equation \ref{Eq:F} assumes that the laser far field is the same as that generated by reflecting a collimated Gaussian beam by a spherical mirror whose optical axis is in the direction of beam propagation. Alternative expressions use $\sqrt{\cos \theta}$ instead of $\cos \theta$ (see Eq. \ref{Eq:F}) to consider the effect of a realistic lens \cite{Tey:09}. I have checked that this difference does not practically affect the reflectivity and transmissivity of the atom array in the geometries used in the simulations.

Using  Eqs. \ref{Eq:E_D}-\ref{Eq:F}, we can write
\begin{equation}
P_{qs}= L_{qs} e^{ikR} a + D_{qs} e^{ikR} \tilde{\sigma}_q,
\label{Eq:P_kq_final}
\end{equation}
where
\begin{eqnarray}
L_{qs} &=& -i \bm{F}(\theta_q,\phi_q) \cdot \bm{\hat{s}}(\theta_q,\phi_q) \sqrt{\frac{\Delta \Omega_q}{A}}   \label{Eq:L_kq} \\ D_{qs} &=& \sqrt{\frac{3\Gamma}{8\pi}} \bm{\hat{u}}_D(\theta_q,\phi_q) \cdot \bm{\hat{s}}(\theta_q,\phi_q) \sqrt{\Delta \Omega_q},
\label{Eq:D_kq}
\end{eqnarray}
where $\tilde{\sigma}_q \equiv \tilde{\sigma}(\theta_q,\phi_q)$ is the collective atom operator for the direction of photodetector $q$ and $\Delta \Omega_q$ is the solid angle covered by photodetector $q$. Since the photodetectors cover the whole surface of a sphere in the far field, $\sum_q \Omega_q = 4 \pi$.

To verify the validity of the method, Appendixes \ref{sec:Master} and \ref{sec:Comparacion} demonstrate that the formulas shown in this section yield the same atomic dynamics as the well-established master equation of collective spontaneous emission, in which the laser is treated classically \cite{Carmichael:00,Clemens:03,Clemens:04}. Although both methods lead to the same atom dynamics, treating the laser as a classical field does not capture the photon statistics in the far field.

\section{Results} \label{sec:Results}

The complete time series of photon detection events of the simulated quantum trajectories provides the necessary information to investigate the directional photon counting statistics. All simulations assume circular polarization, $\hat{\bm u} = \frac{1}{\sqrt{2}} \left( \hat{\bm x} + i \hat{\bm y} \right)$, although linear polarizations would produce the same statistical tendencies. The simulations are carried out using the physical properties of rubidium: $\lambda=780$ nm, and $\Gamma=2 \pi \times 6$ MHz \cite{Volz:96,Mack:11}.

\begin{figure}
\centerline{\scalebox{0.57}{\includegraphics{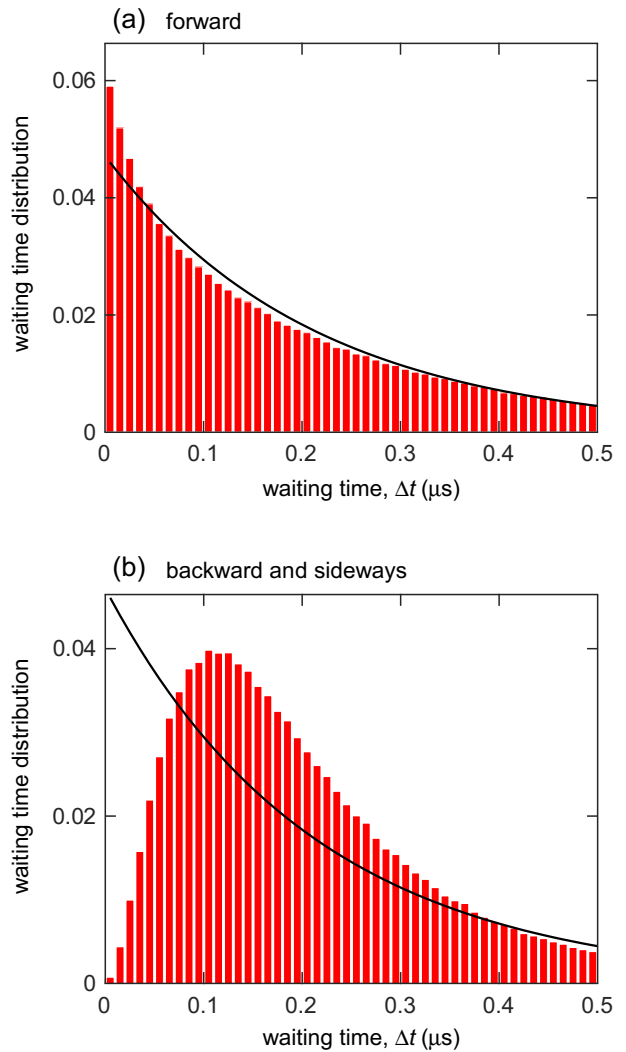}}}
\caption{Normalized waiting-time distributions (a) in the forward direction and (b) in the backward and sideways directions for $N=1$ (red bars). Both plots show the simulations for the Poisson distribution with the same number of photons per unit time (solid black curves). The bin width of both histograms is 10 ns. The Rabi frequency is $\Omega =2 \pi \times 3$ MHz. The beam waist is $w_0=430$ nm ($A_{\text{beam}} = A_{\text{atom}}$).} \label{fig:N1_histograms}
\end{figure}

First, we will study a single atom at the center of the focus, ${\bm r}=(0,0)$. In order to maximize the atom-light coupling, the simulations assume $A_{\text{beam}} = A_{\text{atom}}$, where $A_{\text{atom}} = (3 \lambda^2) / (2 \pi)$ is the absorption cross section of the atom. Notice that in the low-laser-intensity limit the condition $A_{\text{beam}} = A_{\text{atom}}$ does not mean that the atom behaves like a mirror. It means only that the atom scatters all the incoming photons in all directions, including both forward and backward directions. Figure \ref{fig:N1_histograms} shows the distributions of time intervals between two adjacent photons, known as the waiting-time distributions, in the forward direction, and in the backward and sideways directions. For comparison with the classical statistics, both plots include the Poisson distribution corresponding to the same number of photons per unit time. The waiting-time distribution in the forward direction is larger than the Poisson distribution for the smallest waiting times, $\Delta t \lesssim 50$ ns [Fig. \ref{fig:N1_histograms}(a)]. This indicates photon bunching; that is, the transmitted photons are more likely to arrive at the same time than in the classical case. On the contrary, the distribution of the backward and sideways directions shows a dip near zero [Fig. \ref{fig:N1_histograms}(b)]. This indicates photon antibunching; that is, photons tend to arrive with some separation, as expected for a single-photon emitter. Photon antibunching in the backward direction constitutes resonance fluorescence \cite{Lambropoulos:06}, whereas photon bunching in the forward direction is caused by the atom-field correlations of stimulated emission.

\begin{figure}
\centerline{\scalebox{0.57}{\includegraphics{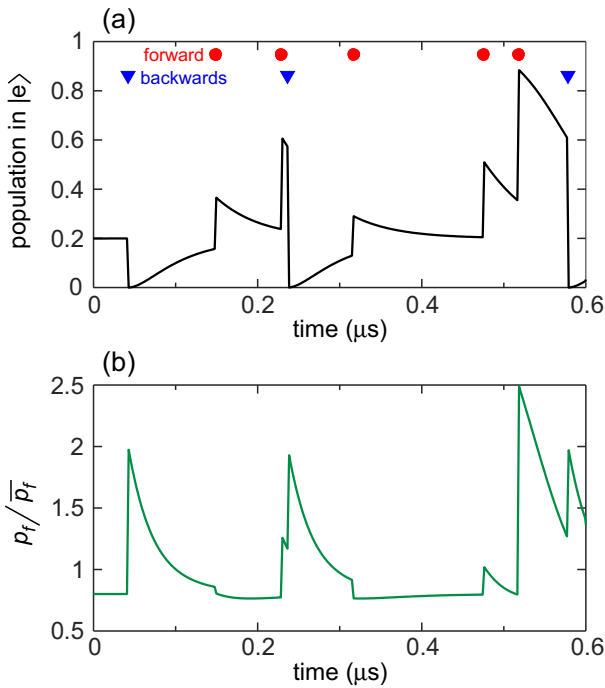}}}
\caption{(a) Time evolution of the excited state population of the atom over a short time interval for $N=1$. Red circles (blue triangles) mark detections of photons in the forward (backward and sideways) direction. (b) Normalized probability of photon detection in the forward direction. The Rabi frequency is $\Omega =2 \pi \times 3$ MHz. The beam waist is $w_0=430$ nm ($A_{\text{beam}} = A_{\text{atom}}$).} \label{fig:Poblaciones_N1}
\end{figure}

\begin{figure}
\centerline{\scalebox{0.3}{\includegraphics{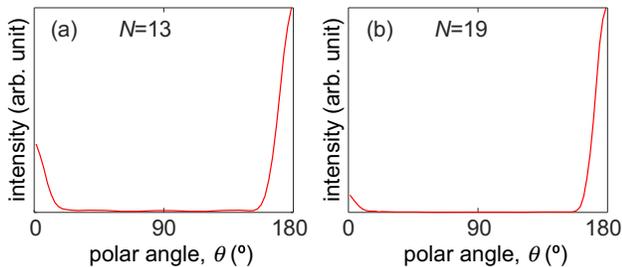}}}
\caption{(a) Radiation intensity for the atom array in Fig. \ref{fig:Figure1}(b) as a function of the polar angle $\theta$. The simulations use $\Omega=2 \pi \times 1$ MHz, $w_0=900$ nm ($A_{\text{beam}} = 4.4 A_{\text{atom}}$), and $d=660$ nm. (b) Radiation intensity for the atom array in Fig. \ref{fig:Figure1}(c). The simulations use $\Omega=2 \pi \times 0.5$ MHz, $w_0=1.1 \mu$m ($A_{\text{beam}} = 6.3 A_{\text{atom}}$), and $d=660$ nm.  } \label{fig:Intensity}
\end{figure}

To gain more insight into the origin of nonclassical photon statistics, let us look at the time evolution of an individual quantum trajectory. Figure \ref{fig:Poblaciones_N1}(a) shows the excited state population over a short time interval. Each time a photon is detected, a quantum jump occurs in the atomic state. Detection of a photon in the backward and sideways directions (blue triangles) projects the atom into its ground state, whereas detection in the forward direction (red circles) usually produces a sudden increase of the excited-state population. The sudden increase in the excited-state population may be surprising since photon detection is usually associated with decay into the ground state. This effect is the consequence of the indistinguishability between the photons of the driving field and the photons of the scattered field, as explained in Ref. \cite{Carmichael:93}. The forward photon flux depends on the phase of the scattered field, which changes randomly over time due to the stochastic nature of the quantum jumps. Its value is given by the phase difference between the excited and ground state coefficients of the wave function. For low laser intensities, its average value is close to $-\pi/2 +\arg{\left( \Omega \right) }$ \cite{Lambropoulos:06}, where the phase of the Rabi frequency is chosen to be $\arg{\left( \Omega \right) }=-\pi$ in all simulations. When a photon is detected in the forward direction, the application of the jump operator $P_f$ on the wave function produces two terms in the ground-state coefficient (see Eq. \ref{Eq:Jump_operators_N1}). These two terms have almost opposite phases, which results in a reduction in the absolute value of the ground-state coefficient and a consequent increase in the excited-state population. In addition, photon detection in the forward direction tends to increase the probability of photon detection in the forward direction, $p_f \equiv \langle \psi(t) | P_f^\dag P_f |\psi (t) \rangle$, as shown in Fig. \ref{fig:Poblaciones_N1}(b). The probability $p_f$ sometimes may also decrease after photon detection if the scattered-field phase is away from its mean value at the moment of detection. Nonetheless, despite the randomness of the quantum jumps, the overall effect is photon bunching in the forward direction.

We are now going to study multiatom systems. Two hexagonal geometries with $N=13$ and $N=19$ are considered, as shown in Figs. \ref{fig:Figure1}(b) and \ref{fig:Figure1}(c). Figure \ref{fig:Intensity} shows the light intensity as a function of the polar angle $\theta$ with respect to the $z$ direction. The reflected power in the 13-atom (19-atom) system is 55\% (82\%) of the incoming power, and the transmitted power is 14\% (5\%). The rest of the power is scattered in the side directions. We consider that transmitted (reflected) photons are those whose polar angle is $\theta < 2 \theta_{\infty}$ ($\theta < \pi-2 \theta_{\infty}$), where $\theta_{\infty}= \lambda / (\pi w_0)$ is the divergence angle of a Gaussian beam. This is $\sim 16^{\circ}$ ($\sim 13^{\circ}$) in the simulations with $N=13$ ($N=19$).

\begin{figure}
\centerline{\scalebox{0.57}{\includegraphics{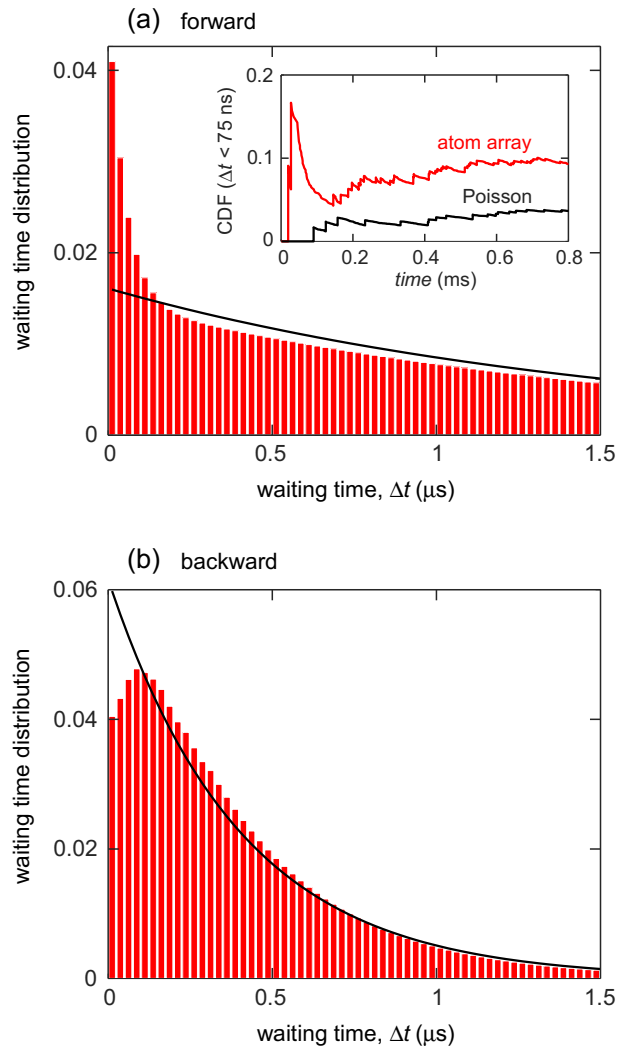}}}
\caption{Normalized waiting-time distributions of the (a) transmitted and (b) reflected beams for the atom array in Fig. \ref{fig:Figure1}(b) with $N=13$ (red bars). The bin width of both histograms is 25 ns. The simulations use $\Omega=2 \pi \times 1$ MHz ($|\alpha|^2=4.6 \times 10^6$ s$^{-1}$), $w_0=900$ nm ($A_{\text{beam}} = 4.4 A_{\text{atom}}$), and $d=660$ nm $(d \simeq 0.85 \lambda)$. Inset: cumulative distribution functions (CDFs) for $\Delta t \lesssim$ 75 ns, which represents the number of photon pairs divided by the total photon number in the forward direction, as a function of time for a particular quantum trajectory (red). Both plots show the simulations for the Poisson distribution with the same number of photons per unit time (solid black curves).} \label{fig:N13_histograms}
\end{figure}

\begin{figure}
\centerline{\scalebox{0.57}{\includegraphics{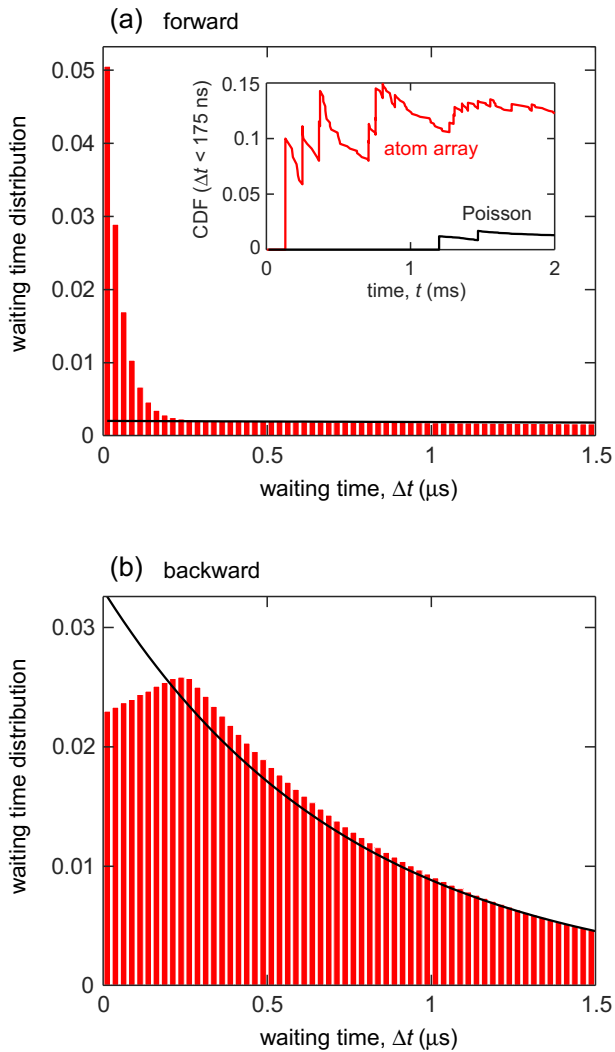}}}
\caption{(a) Normalized waiting-time distributions of the (a) transmitted and (b) reflected beams for the atom array in Fig. \ref{fig:Figure1}(c) with $N=19$ (red bars). The bin width of both histograms is 25 ns. The simulations use $\Omega=2 \pi \times 0.5$ MHz ($|\alpha|^2=1.6 \times 10^6$ s$^{-1}$), $w_0=1.1 \mu$m ($A_{\text{beam}} = 6.3 A_{\text{atom}}$), and $d=660$ nm $(d \simeq 0.85 \lambda)$. Inset: CDFs for $\Delta t \lesssim$ 175 ns, which represents the number of photon pairs divided by the total photon number in the forward direction,  as a function of time for a particular quantum trajectory (red). Both plots show the simulations for the Poisson distribution with the same number of photons per unit time (solid black curves).} \label{fig:N19_histograms}
\end{figure}

Figures \ref{fig:N13_histograms} and \ref{fig:N19_histograms} show the normalized waiting time distributions for $N=13$ and $N=19$, respectively. Like in the single-atom system, the transmitted photons are bunched and the reflected photons are antibunched. The deviations from Poisson statistics are more pronounced for smaller waiting times. In the transmitted beam for $N=13$ $(N=19)$ the probability that $\Delta t \lesssim$ 75 ns ($\Delta t \lesssim$ 175 ns) is $\simeq 0.1$ ($\simeq 0.12$), while it is only $\simeq 0.05$ ($\simeq 0.01$) in the Poisson distribution. This means that up to $\simeq 20$ \% ($\simeq 25$ \%) of transmitted photons are grouped in pairs. Interestingly, the largest deviations from classical statistics are found for $N=19$.

Almost all bunches are formed by photon pairs. Bunches of three or more photons are unlikely for the laser intensities considered in these simulations. They become more likely for higher laser intensities. However, increasing the laser intensity also leads to smaller temporal separations between bunches, thus producing an increasingly Poissonian statistics.

The question now is whether deviations from the Poisson statistics can be observed in an experimentally feasible time. Building a histogram with very narrow bins can take too long under certain experimental conditions. Fortunately, the evaluation of the number of photon pairs does not require obtaining the complete histogram. To know the number of photon pairs in the forward direction, it is sufficient to count the number of waiting time intervals below a certain value. We consider that there is a photon pair when the time interval is $\Delta t \lesssim$ 75 ns for the 13-atom system and $\Delta t \lesssim$ 175 ns for the 19-atom system. These values correspond to higher waiting time probabilities than in the Poisson distribution in the histograms in Figs. \ref{fig:N13_histograms}(a) and \ref{fig:N19_histograms}(a), respectively. The insets in Figs. \ref{fig:N13_histograms}(a) and \ref{fig:N19_histograms}(a) plot the number of photon pairs divided by the number of photons detected in the forward direction as a function of time for a particular quantum trajectory. These functions are also known as the cumulative distribution functions (CDF) because they integrate over all waiting times $\Delta t$ below a certain value. As we can see, it takes only a few hundred microseconds for the deviations from Poisson statistics to become clearly observable. During this time each atom scatters $\lesssim 70$ photons. A two-dimensional atomic array that is robust against such a number of scattered photons is possible, as demonstrated in recent experiments \cite{Rui:20}. In the example with $N=19$, after random initial dynamics, the CDF tends to the constant value 0.13, which means that up to 25\% of the transmitted photons are grouped in pairs.

\begin{figure}
\centerline{\scalebox{0.57}{\includegraphics{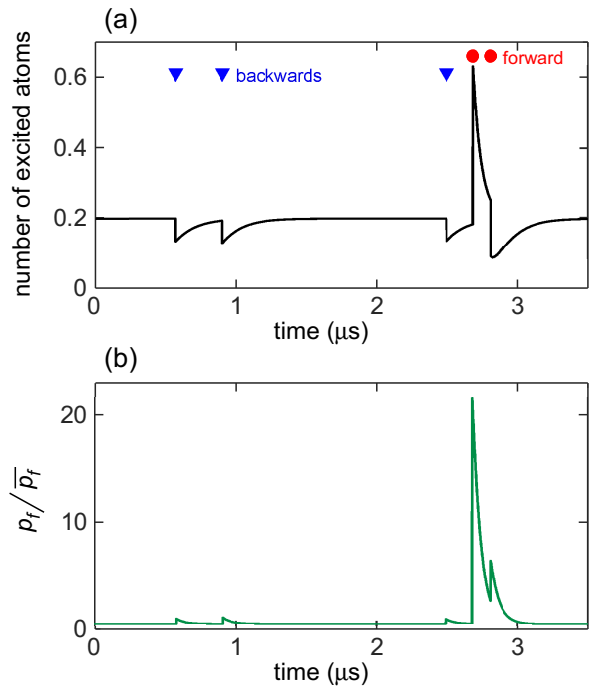}}}
\caption{(a) Time evolution of the number of excited atoms, $\sum_{n=1}^{N} \langle \sigma _n^\dag \sigma _n \rangle$, over a short time interval of an individual quantum trajectory for $N=19$. Blue triangles (red circles) mark detections of photons in the backward (forward) direction. (b) Normalized probability of photon detection in the forward direction. The parameters are the same as in Fig. \ref{fig:N19_histograms}} \label{fig:Poblaciones_N19}
\end{figure}

Figure \ref{fig:Poblaciones_N19}(a) shows the number of excited atoms for an individual quantum trajectory, $ \sum_{n=1}^{N} \langle \sigma _n^\dag \sigma _n \rangle$, where $\sigma_n \equiv |g_n\rangle \langle e_n|$ is the operator that lowers the state of atom $n$. Figure \ref{fig:Poblaciones_N19}(b) shows the photon transmission probability, $p_f$, divided by its time average. The effects are similar to those for the single-atom system shown in Fig. \ref{fig:Poblaciones_N1}. Detection of a photon in the backward direction removes one excitation from the atomic array, whereas detection of a photon in the forward direction can produce a sudden increase in the excited state population, depending on the phase of the atomic dipole moment at the time of photon detection. The overall effect is photon bunching in the transmitted beam. The arguments used for the one-atom system are also valid for the atomic arrays, except that here the scattered field is the superposition of the individual fields of all the atoms and the jump operator is $P_{qs}$ (see Eq. \ref{Eq:P_kq_final}).

Strikingly, as the number of atoms increases, photon bunching in the transmitted beam does not disappear. This is due to the cooperative nature of the interactions between the atoms and the Gaussian beam mode. When a photon is detected in the forward direction, the whole ensemble is projected onto a state that has lower reflectivity and higher population in the excited state. This increases the probability of a second photon count. Something different happens with the reflected light. As the number of atoms increases, the photon statistics of the reflected beam approaches the Poisson distribution. The reason is that multiple excitations in the atomic array are not forbidden, and the probability of simultaneous reflection of more than one photon is not zero.

\section{Summary} \label{sec:discussion}

This paper has described a quantum trajectory method to simulate the photon statistics with angular resolution of an atomic array illuminated by a laser field. The most interesting effect occurs in the transmitted beam, where the photons of the laser are indistinguishable from the photons reemitted by the atoms in the direction of beam propagation. The quantum interference between the laser field and the dipolar field of the atoms gives rise to photon bunching in the transmitted beam. This photon statistics is a signature of the cooperative quantum nature of atom-light interactions in atom arrays with sub-wavelength spacing. In the example with $N=19$, up to 25\% of the transmitted photons are grouped in pairs. The proposed measurements are experimentally feasible using the two-dimensional atomic arrays realized in recent works \cite{Rui:20}.

\vspace{3mm}
\appendix

\section{Master equation for the internal dynamics of the atoms} \label{sec:Master}

The quantum trajectory algorithm used in the simulations is based on the master equation of collective spontaneous emission \cite{Carmichael:00,Clemens:03,Clemens:04} with a quantum description of the laser field incorporated into it. On the other hand, we know that the atomic dynamics can be simulated by describing the laser field classically. In fact, the only reason to have described the laser as a quantum field is to take into account the indistinguishability between the laser photons and the forward scattered photons, as this is needed to simulate the directional photon statistics in the far field. This appendix will show that the commonly used master equation of collective spontaneous emission \cite{Carmichael:00,Clemens:03,Clemens:04}, in which the laser field is described classically, can be derived from equations in Sec. \ref{sec:Formulation}, where the laser is described as a quantum field. This appendix will also show that the Hamiltonian ${\cal H}_{AF}$ in Eq. \ref{Eq:H_AF}, which has the form of an atom-cavity interaction Hamiltonian, can be transformed into the atom-laser interaction Hamiltonian described in most quantum optics books without quantum field operators \cite{Lambropoulos:06}.

First, we write the master equation of collective spontaneous emission in which the laser field is described classically \cite{Lambropoulos:06,Clemens:03},
\begin{equation}
\frac{\partial}{\partial t} \rho_A = -\frac{i}{\hbar} \left[ {\cal H}_{\text{eff}},\rho_A \right] + {\cal L}_{A} \rho_A,
\label{Eq:Master_atoms}
\end{equation}
where $\rho_A$ is the density operator of the atom array; ${\cal H}_{\text{eff}}$ is the effective Hamiltonian of the atom-field interactions,
\begin{equation}
{\cal H}_{\text{eff}} = \hbar \Omega \sum_{n=1}^{N} \left( f_n \sigma_n^\dag + f_n^\ast \sigma_n \right),
\label{Eq:H_eff}
\end{equation}
and ${\cal L}_{A}$ is the Lindblad operator of the collective atomic decay,
\begin{equation}
{\cal L}_{A} \rho = \sum_{n \neq m=1}^{N} \frac{\Gamma_{nm}}{2} \left( 2 \sigma_n  \rho \sigma_m^\dag - \sigma_m^\dag \sigma_n \rho -  \rho \sigma_m^\dag \sigma_n \right), \label{Eq:Lindblad_A_2}
\end{equation}
Here, $\Gamma_{nm}$ are the coefficients of spontaneous photon emission \cite{Carmichael:00,Clemens:03},
\begin{eqnarray}
\Gamma_{nm}  & = & \frac{3 \Gamma}{2} \left[ \left( 1-|\bm{ u} \cdot \bm{\hat{r}_{nm}}|^2\right) \frac{\sin \xi_{nm}}{\xi_{nm}} \right. \nonumber \\
 & + & \left. \left( 1-3 |\bm{ u} \cdot \bm{\hat{r}_{nm}}|^2\right) \left(\frac{\cos \xi_{nm}}{\xi_{nm}^2}-\frac{\sin \xi_{nm}}{\xi_{nm}^3} \right) \right], \label{Eq:Gamma_nm}
\end{eqnarray}
where $\xi_{nm}\equiv 2 \pi r_{nm}/\lambda$.

In what follows, this appendix will demonstrate that Eqs. \ref{Eq:Master_atoms}-\ref{Eq:Gamma_nm} can be derived from the equations in the main text. The Sch\"{o}dinger equation in Eq. \ref{Eq:Schroedinger} corresponds to the following master equation \cite{Lambropoulos:06},
\begin{equation}
\frac{\partial}{\partial t} \rho = -\frac{i}{\hbar} \left[ {\cal H}_{AF} + {\cal H}_{AA},\rho \right] + {\cal L} \rho,
\label{Eq:Master}
\end{equation}
where $\rho$ is the density operator of the atom-field system, and ${\cal L}$ is the Lindblad operator,
\begin{equation}
{\cal L} \rho \equiv \sum_{q=1}^Q \sum_{s} \left( P_{qs} \rho P^\dag_{qs} + \frac{1}{2} P^\dag_{qs} P_{qs} \rho + \frac{1}{2} \rho P^\dag_{qs} P_{qs} \right). \label{Eq:Lindblad_1}
\end{equation}
The operator ${\cal L}$ can be separated into three parts,
\begin{equation}
{\cal L} \rho \equiv \left( {\cal L}_F + {\cal L}_A + {\cal L}_{AF} \right) \rho , \label{Eq:Lindblad_3parts}
\end{equation}
where ${\cal L}_F$ contains the terms $|L_{qs}|^2$ with the products of field operators, ${\cal L}_A$ contains the terms $|D_{qs}|^2$ with the products of collective atomic operators, and ${\cal L}_{AF}$ contains the crossed terms $L^\ast_{qs} D_{qs}$ and $L_{qs} D^\ast_{qs}$ with the products of field and atomic operators. In the following sections, the operators ${\cal L}_F$, ${\cal L}_A$, and ${\cal L}_{AF}$ will be transformed into convenient mathematical expressions that allow us to derive Eqs. \ref{Eq:Master_atoms}-\ref{Eq:Gamma_nm} from the equations in Secs. \ref{sec:Algorithm} and \ref{sec:Formulation}. The procedure consists of summing over all directions and polarizations to arrive at simplified expressions for the operators.

\subsection{Operator with the field terms, ${\cal L}_{F}$} \label{sec:L_F}

First, we prove that ${\cal L}_F$ takes the form of the Lindblad operator of an open field. Using Eqs. \ref{Eq:P_kq_final}-\ref{Eq:D_kq}, \ref{Eq:Lindblad_1}, and \ref{Eq:Lindblad_3parts}, we find
\begin{eqnarray}
{\cal L}_{F} \rho &=& \sum_{q=1}^Q \sum_{s} \frac{|L_{qs}|^2}{2} \left( 2 a \rho a^\dag - a^\dag a \rho -  \rho a^\dag a \right). \label{Eq:Lindblad_F_1}
\end{eqnarray}
In order to sum over all directions and polarizations, we use the following identity, which can be verified by finite-element integration,
\begin{eqnarray}
\sum_{q=1}^Q \sum_{s} | \bm{F}^\ast(\theta_q,\phi_q)|^2 \Delta \Omega_q = A_{\text{beam}}. \label{Eq:sum_F}
\end{eqnarray}
Note that the result in Eq. \ref{Eq:sum_F} is a necessary condition for the conservation of the electromagnetic energy. We now can write
\begin{equation}
\sum_{q=1}^Q \sum_{s} |L_{qs}|^2 = -i.
\label{Eq:sum_L_kq}
\end{equation}
Using Eqs. \ref{Eq:Lindblad_F_1} and \ref{Eq:sum_L_kq}, we obtain the desired result,
\begin{eqnarray}
{\cal L}_{F} \rho &=& - \frac{i}{2} \left( 2 a \rho a^\dag - a^\dag a \rho -  \rho a^\dag a \right). \label{Eq:Lindblad_F_2}
\end{eqnarray}

\subsection{Operator with the atomic terms, ${\cal L}_{A}$} \label{sec:L_A}

The expression for ${\cal L}_{A}$ can be obtained by combining Eqs. \ref{Eq:P_kq_final}-\ref{Eq:D_kq}, \ref{Eq:Lindblad_1}, and \ref{Eq:Lindblad_3parts}. We find
\begin{equation}
{\cal L}_{A} \rho = \sum_{q=1}^Q \sum_{s} \frac{|D_{qs}|^2}{2} \left( 2 \tilde{\sigma}_{q}  \rho \tilde{\sigma}_{q}^\dag - \tilde{\sigma}_{q}^\dag \tilde{\sigma}_{q} \rho -  \rho \tilde{\sigma}_{q}^\dag \tilde{\sigma}_{q} \right), \label{Eq:Lindblad_A}
\end{equation}
Next, Eq. \ref{Eq:Lindblad_A} must be transformed into an expression with single-atom operators, using Eq. \ref{Eq:collective_sigma}. For this, we first verify the following equation by finite-element integration,
\begin{equation}
\sum_{q=1}^Q \sum_{s} e^{-ik \left( \bm{\hat{r}_n} - \bm{\hat{r}_m} \right) \cdot  \bm{\hat{R}_q}} |D_{qs}|^2 = \Gamma_{nm}, \label{Eq:sum_Gamma_nm}
\end{equation}
Finally, by combining \ref{Eq:Lindblad_A} and \ref{Eq:sum_Gamma_nm}, we arrive at Eq. \ref{Eq:Lindblad_A_2}.

\subsection{Operator with the cross terms, ${\cal L}_{AF}$} \label{sec:discussion}

The Lindblad operator ${\cal L}_{AF}$ contains the cross terms $L^\ast_{qs} D_{qs}$ and $L_{qs} D^\ast_{qs}$. To express ${\cal L}_{AF}$ as a function of single-atom operators, we use
\begin{equation}
\sum_{q=1}^Q \sum_{s} L^\ast_{qs} D_{qs} = -i g \sum_{n=1}^N  f_n^\ast,
\label{Eq:sum_LD}
\end{equation}
which is obtained from
\begin{equation}
\sum_{q=1}^Q \sum_{s} \bm{F}^\ast(\theta_q,\phi_q) \cdot \bm{\hat{u}}_D(\theta_q,\phi_q) \Delta \Omega_q e^{-ik \bm{\hat{r}_n} \cdot  \bm{\hat{R}_q}} = \lambda f_n^\ast,
\label{Eq:sum_F_uD}
\end{equation}
where $\bm{\hat{R}_q}=(\sin \theta_q \cos \phi_q, \cos \theta_q \cos \phi_q, -\sin \phi_q)$. Equation \ref{Eq:sum_F_uD} was checked by finite-element numerical integration. In this way, we find
\begin{eqnarray}
{\cal L}_{AF} \rho &=& \frac{i}{2}g \sum_{n=1}^N f_n \left(2 a \rho \sigma_n^\dag  - \sigma_n^\dag a \rho - \rho \sigma_n^\dag a   \right) \nonumber \\
& - & \frac{i}{2}g \sum_{n=1}^N f_n^\ast \left(2 \sigma_n \rho a^\dag  - a^\dag \sigma_n \rho - \rho a^\dag \sigma_n   \right). \label{Eq:Lindblad_AF}
\end{eqnarray}

\subsection{Elimination of the field operators} \label{sec:Master_without_field_operators}

The final step of the demonstration consists of inserting the expressions for ${\cal L}_{F}$, ${\cal L}_{A}$, and ${\cal L}_{AF}$ into Eqs. \ref{Eq:Master}-\ref{Eq:Lindblad_3parts}. Since the field is in the classical state $|\alpha\rangle$, the following substitutions are made, $a \rightarrow \alpha$ and $a^\dag \rightarrow \alpha^\ast$. Using $\Omega=2 g \alpha$, we arrive at Eqs. \ref{Eq:Master_atoms}-\ref{Eq:Gamma_nm}, as desired.

\section{Comparison between different methods} \label{sec:Comparacion}

\begin{figure}
\centerline{\scalebox{0.57}{\includegraphics{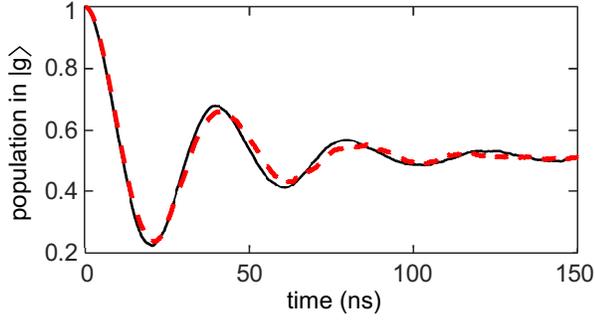}}}
\caption{Time evolution of the ground-state population for a single atom located at the center of the focus. The Rabi frequency is $\Omega=2 \pi \times 25$ MHz. Red dashed line: average over 5000 trajectories using the quantum trajectory method described in Secs. \ref{sec:Algorithm} and \ref{sec:Formulation} ($w_0=1.2 \lambda$). Solid black line: average over 5000 trajectories using a quantum trajectory method based on Eqs. \ref{Eq:Master_atoms}-\ref{Eq:Gamma_nm}, as explained in Refs. \cite{Masson:20,Clemens:03}.} \label{fig:FigureC}
\end{figure}

\begin{figure}
\centerline{\scalebox{0.57}{\includegraphics{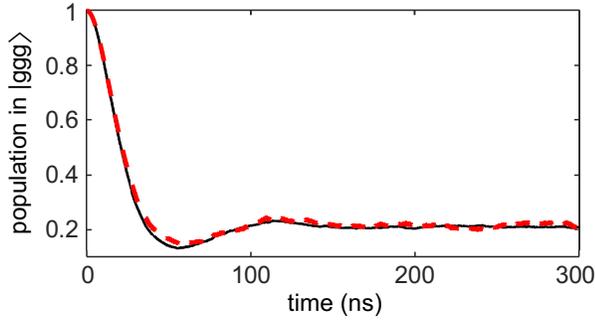}}}
\caption{Time evolution of the triple ground-state population for a three-atom system with $\bm r_1 = (0,0)$, $\bm r_2 = (d,0)$, and $\bm r_3 = (0,0.7d)$, where $d=0.6 \lambda$. The Rabi frequency is $\Omega=2 \pi \times 10$ MHz. Red dashed line: average over 1000 trajectories using the quantum trajectory method described in Secs. \ref{sec:Algorithm} and \ref{sec:Formulation} ($w_0=1.2 \lambda$). Solid black line: average over over 5000 trajectories using a quantum trajectory method based on Eqs. \ref{Eq:Master_atoms}-\ref{Eq:Gamma_nm}, as explained in Refs. \cite{Masson:20,Clemens:03}.} \label{fig:FigureB}
\end{figure}

\begin{figure}
\centerline{\scalebox{0.57}{\includegraphics{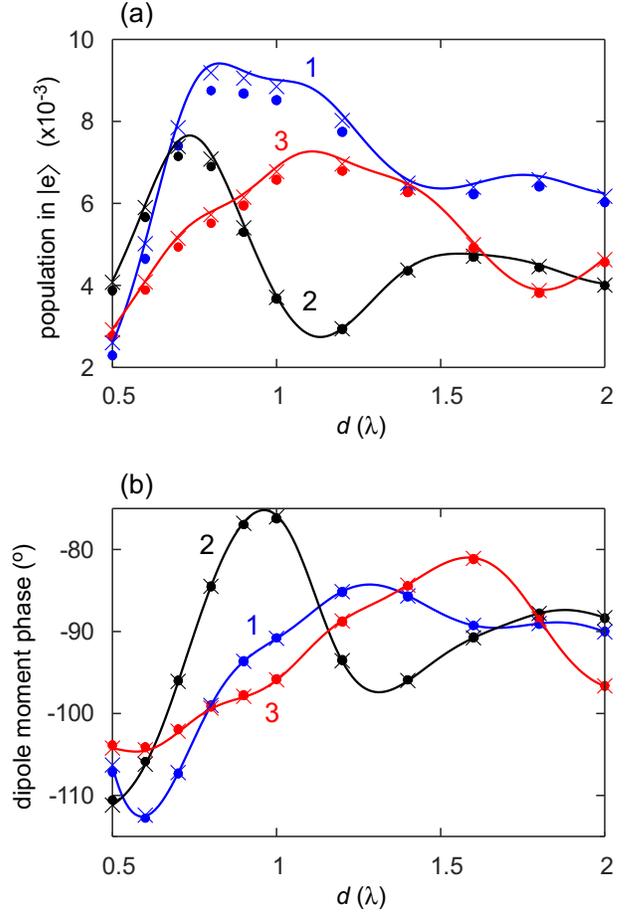}}}
\caption{(a) Excited-state populations and (b) dipole moment phases in a three-atom system in the low-laser-intensity limit as a function of the atom-atom separation. The positions are $\bm r_1 = (0,0)$, $\bm r_2 = (d,0)$, and $\bm r_3 = (0,0.7d)$, where $d$ varies from $0.5 \lambda$ to $2 \lambda$ and $w_0=2d$. Solid lines: calculations using the method of Refs. \cite{Bettles:15,Bettles:16}, which assumes that the atoms are classical dipoles. Circles: calculations with the quantum trajectory method described in Secs. \ref{sec:Algorithm} and \ref{sec:Formulation}. Crosses: calculations using a quantum trajectory method based on Eqs. \ref{Eq:Master_atoms}-\ref{Eq:Gamma_nm} (Refs. \cite{Masson:20,Clemens:03}).} \label{fig:FigureA}
\end{figure}

This appendix compares the numerical solutions obtained with different methods in order to check that the calculations have been performed correctly. Figures \ref{fig:FigureC} and \ref{fig:FigureB} show the ground state populations calculated for two particular cases, with $N=1$ and $N=3$. Calculations were carried out by means of quantum trajectories, either using the equations in Secs. \ref{sec:Algorithm} and \ref{sec:Formulation} (red dashed lines) or using Eqs. \ref{Eq:Master_atoms}-\ref{Eq:Gamma_nm} (black solid lines). Both methods give practically the same solution, which corroborates the validity of our algorithm. The very small differences between the two methods are most likely due to the fact that Eq. \ref{Eq:F} is an approximation of the far field.

The quantum trajectories of Eqs. \ref{Eq:Master_atoms}-\ref{Eq:Gamma_nm} are calculated using source-mode jump operators. Here, I do not give the details of how to obtain the source-mode jump operators because this has been exhaustively described elsewhere \cite{Masson:20,Clemens:03}. In a few words, the set of $N$ source-mode jump operators is obtained by diagonalizing the matrix of coefficients $\Gamma_{nm}$ in Eq. \ref{Eq:Gamma_nm}. Unlike the jump operators $P_{qs}$, source-mode jump operators cannot be identified with a particular photon detection event.

Figure \ref{fig:FigureA} shows the excited state populations $\langle \sigma _n^\dag \sigma _n \rangle$ and the dipole moment phases $\text{arg}\{\langle \sigma _n \rangle\}$ of a three-atom system in a classical field ($n=1,2,3$). The results obtained with three different methods are compared with each other: the quantum trajectory method from Sec. \ref{sec:Algorithm}, the quantum trajectory method using Eqs. \ref{Eq:Master_atoms}-\ref{Eq:Gamma_nm}, and the classical method of Refs. \cite{Bettles:15,Bettles:16}, which is valid in the low-laser-intensity limit.


\begin{thebibliography}{10}



\bibitem{Yu:14} N. Yu and F. Capasso, Nat. Mater. \textbf{13}, 139 (2014).
\bibitem{Chen:16} H.-T. Chen, A.~J. Taylor, and N. Yu, Rep. Prog. Phys. \textbf{79}, 076401 (2016).

\bibitem{Stav:18} T. Stav, A. Faerman, E. Maguid, D. Oren, V. Kleiner, E. Hasman, and M. Segev, Science \textbf{361}, 1101-1104 (2018).
\bibitem{Wang:18} K. Wang, J.~G. Titchener, S.~S. Kruk, L.~Xu, H.-P. Chung, M. Parry, I.~I. Kravchenko, Y.-H. Chen, A.~S. Solntsev, Y.~S. Kivshar, D.~N. Neshev, and A.~A. Sukhorukov, Science \textbf{361}, 1104-1108 (2018).
\bibitem{Lyons:19} A. Lyons, D. Oren, T. Roger, V. Savinov, J. Valente, S. Vezzoli, N.~I. Zheludev, M. Segev, and D. Faccio, Phys. Rev. A \textbf{99}, 011801(R) (2019).


\bibitem{Rui:20} J. Rui, D. Wei, A. Rubio-Abadal, S. Hollerith, J. Zeiher, D.~M. Stamper-Kurn, C. Gross, and I. Bloch, Nature \textbf{583}, 369–374 (2020).

\bibitem{Bekenstein:20} R. Bekenstein, I. Pikovski, H. Pichler, E. Shahmoon, S.~F. Yelin, and M.~D. Lukin, Nat. Phys. \textbf{16}, 676–681 (2020).


\bibitem{Bettles:15} R.~J. Bettles, S.~A. Gardiner, and C.~S. Adams, Phys. Rev. A \textbf{92}, 063822 (2015).
\bibitem{Bettles:16} R.~J. Bettles, S.~A. Gardiner, and C.~S. Adams, Phys. Rev. Lett. \textbf{116}, 103602 (2016).
\bibitem{Shahmoon:17} E. Shahmoon, D.~S. Wild, M.~D. Lukin, and S.~F. Yelin, Phys. Rev. Lett. \textbf{118}, 113601 (2017).
\bibitem{Yoo:20} S.-M. Yoo and J. Javanainen, Opt. Express \textbf{28}, 9764-9776 (2020).


\bibitem{Parmee:20} C.~D. Parmee and J. Ruostekoski, Commun. Phys. \textbf{3}, 205 (2020).
\bibitem{Parmee:21} C.~D. Parmee and J. Ruostekoski, Phys. Rev. A \textbf{103}, 033706 (2021).
\bibitem{Bettles:20} R.~J. Bettles, M.~D. Lee, S.~A. Gardiner, and J. Ruostekoski, Commun. Phys. \textbf{3}, 141 (2020).

\bibitem{Brandes:05} T. Brandes, Phys. Rep. \textbf{408}, 315–474 (2005).
\bibitem{Guerin:17} W. Guerin, M. Rouabah, and R. Kaiser, J. Mod. Opt. \textbf{64}, 895–907 (2017).


\bibitem{Svidzinsky:10} A.~A. Svidzinsky, J.-T. Chang, and M.~O. Scully, Phys. Rev. A \textbf{81}, 053821 (2010).
\bibitem{Feng:13} W. Feng, Y. Li, and S.-Y. Zhu, Phys. Rev. A \textbf{88}, 033856 (2013).
\bibitem{Feng:14} W. Feng, Y. Li, and S.-Y. Zhu, Phys. Rev. A \textbf{89}, 013816 (2014).
\bibitem{Facchinetti:16} G. Facchinetti, S.~D. Jenkins, and J. Ruostekoski, Phys. Rev. Lett. \textbf{117}, 243601 (2016).
\bibitem{Arruda:20} T.~J. Arruda, R. Bachelard, J. Weiner, S. Slama, and P.~W. Courteille, Phys. Rev. A \textbf{101}, 023828 (2020).
\bibitem{Zhang:20} Y.-X. Zhang and K. M\o{}lmer, Phys. Rev. Lett. \textbf{125}, 253601 (2020).
\bibitem{Alaee:20} R. Alaee, A. Safari, V. Sandoghdar, and R.~W. Boyd, Phys. Rev. Research 2, 043409 (2020).
\bibitem{Petrosyan:21} D. Petrosyan and K. M\o{}lmer, Phys. Rev. A \textbf{103}, 023703 (2021).
\bibitem{Gulfam:18} Q.-u.-A. Gulfam and Z. Ficek, Phys. Rev. A \textbf{98}, 063824 (2018).
\bibitem{Masson:20} S.~J. Masson, I. Ferrier-Barbut, L.~A. Orozco, A. Browaeys, and A. Asenjo-Garcia, Phys. Rev. Lett. \textbf{125}, 263601 (2020). 

\bibitem{Bhatti:15} D. Bhatti, J. von Zanthier, and G.~S. Agarwal, Sci. Rep. \textbf{5}, 17335 (2015).
\bibitem{Liberal:19} I. Liberal, I. Ederra, and R.~W. Ziolkowski, Photonics \textbf{6}, 14 (2019).
\bibitem{Williamson:20} L. A. Williamson, M. O. Borgh, and J. Ruostekoski, Phys. Rev. Lett. \textbf{125}, 073602 (2020).
\bibitem{Holzinger:21} R. Holzinger, M. Moreno-Cardoner, and H. Ritsch, Appl. Phys. Lett. \textbf{119}, 024002 (2021).


\bibitem{Einstein:16} A. Einstein, Verhandlungen der Deutschen Physikalischen Gesellschaft \textbf{18}, 318–323 (1916).

\bibitem{Cray:82} M. Cray, M.‐L. Shih, and P.~W. Milonni, Am. J. Phys. \textbf{50}, 1016 (1982).

\bibitem{Tey:08} M.~K. Tey, Z. Chen, S.~A. Aljunid, B. Chng, F. Huber, G. Maslennikov, and C. Kurtsiefer, Nat. Phys. \textbf{4}, 924-927 (2008).
\bibitem{Tey:09} M.~K. Tey, G. Maslennikov, T.~C.~H. Liew, S.~A. Aljunid, F. Huber, B. Chng, Z. Chen, V. Scarani, and C. Kurtsiefer, New J. Phys. \textbf{11}, 043011 (2009).

\bibitem{Carmichael:00} H.~J. Carmichael and K. Kim, Opt. Commun. \textbf{179} 417–427 (2000).
\bibitem{Clemens:03} J.~P. Clemens, L. Horvath, B.~C. Sanders, and H.~J. Carmichael, Phys. Rev. A \textbf{68}, 023809 (2003).
\bibitem{Clemens:04} J.~P. Clemens, L. Horvath, B.~C. Sanders, and H.~J. Carmichael, J. Opt. B: Quantum Semiclass. Opt. \textbf{6}, S736-S741 (2004).

\bibitem{Carmichael:93} H.~J. Carmichael, Phys. Rev. Lett. \textbf{70}, 2273 (1993).
\bibitem{Gardiner:94} C.~W. Gardiner and A.~S. Parkins, Phys. Rev. A \textbf{50}, 1792 (1994).
\bibitem{Nha:05} H. Nha and H.~J. Carmichael, Phys. Rev. A \textbf{71}, 013805 (2005).
\bibitem{Noh:08} C. Noh and H.~J. Carmichael, Phys. Rev. Lett. \textbf{100}, 120405 (2008).
\bibitem{Zhang:18} X.~H.~H. Zhang and H.~U. Baranger, Phys. Rev. A \textbf{97}, 023813 (2018).

\bibitem{Dalibard:92} J. Dalibard, Y. Castin, and K. M\o{}lmer, Phys. Rev. Lett. \textbf{68}, 580 (1992).
\bibitem{Molmer:96} K. M\o{}lmer and Y. Castin, Quantum Semiclass. Opt. \textbf{8}  49–72 (1996).
\bibitem{Lambropoulos:06} P. Lambropoulos and D. Petrosyan, \textit{Fundamentals of Quantum Optics and Quantum Information} (Springer, Berlin, 2006). 



\bibitem{Volz:96} U. Volz and H. Schmoranzer, Phys. Scr. \textbf{65}, 48-56 (1996).
\bibitem{Mack:11} M. Mack, F. Karlewski, H. Hattermann, S. H\"{o}ckh, F. Jessen, D. Cano, J. Fort\'{a}gh, Phys. Rev. A \textbf{83}, 052515 (2011).


\bibitem{Jackson:67} J.~D. Jackson , \textit{Classical electrodynamics} (John Wiley $\&$ Sons, Inc., 1967).



\end{thebibliography}
\end{document}